\documentclass[journal=jacsat,manuscript=article]{achemso}

\usepackage{chemformula} 
\usepackage[T1]{fontenc} 
\usepackage{xcolor}
\usepackage{physics}
\usepackage{amsmath,bm}
\usepackage{amsmath,MnSymbol}
\usepackage{amsmath}

\author{Abhineet Singh Rajput}
\alsoaffiliation{$^{\dagger}$Department of Mechanical Engineering, Indian Institute of Science, Bangalore, India}
\author{Sarath Chandra Varma}
\alsoaffiliation{$^{\dagger}$Department of Mechanical Engineering, Indian Institute of Science, Bangalore, India}
\author{Aloke Kumar}
\alsoaffiliation{$^{\dagger}$Department of Mechanical Engineering, Indian Institute of Science, Bangalore, India}
\email{*alokekumar@iisc.ac.in}

\title[An \textsf{achemso} demo]
  {Sub-Newtonian coalescence in polymeric fluids}

\abbreviations{IR,NMR,UV}

\begin{document}







\begin{abstract}
We present a theoretical framework for capturing the coalescence of a pendant drop with a sessile drop in polymeric fluids. The framework is based on the unification of various constitutive laws under high Weissenberg creeping flow limit. Our results suggests that the phenomenon comes under a new regime namely, the sub-Newtonian regime followed by the limiting case of arrested coalescence with the arrest angle $\theta_{arrest}\propto Ec^{-1/2}$, where $Ec$ is the Elasto-capillary number. Further, we propose a new time scale $T^*$ integrating the continuum variable $Ec$ and the macromolecular parameter $N_e$, the entanglement density to describe the liquid neck evolution. Finally, we validate the framework with high speed imaging experiments performed across different molecular weights of Poly(ethylene oxide) (PEO).
\end{abstract}

\section{Introduction}
Coalescence is an energy minimization phenomenon in which two drops merge to form a thermodynamically stable daughter drop\cite{76}. Coalescence of droplets of Newtonian fluids plays a key role in rain drop condensation\cite{1,3}, combustion\cite{14}, atomization of metal droplets\cite{HOPFES2021103723}; while non-Newtonian fluid droplets coalescence finds applications in food industry\cite{13}, spray coating and paintings\cite{5,6}, even processes linked to life like those in growth and development of tumor\cite{tumor}. Despite the varied and versatile application of non-Newtonian fluids, coalescence dynamics of such fluids remains a sparsely studied area. The vastness of the domain of non-Newtonian fluids - they can range from macromolecular fluids to various colloids - makes a unified understanding even more elusive. Each subclass has a different micro-structure composition leading to distinct behaviors. However, there are few recent studies on a special class of non-Newtonian fluids i.e macromolecular fluids\cite{our,varma2021coalescence,varma2022rheocoalescence,varma2022elasticity,CHEN2022283} that have highlighted the deviation from proposed Newtonian behaviour. But a generalized theoretical framework unifying various constitutive laws to probe the phenomenon in viscoelastic fluids remains an open question.

\begin{figure*}[h!]
\includegraphics[width=\linewidth]{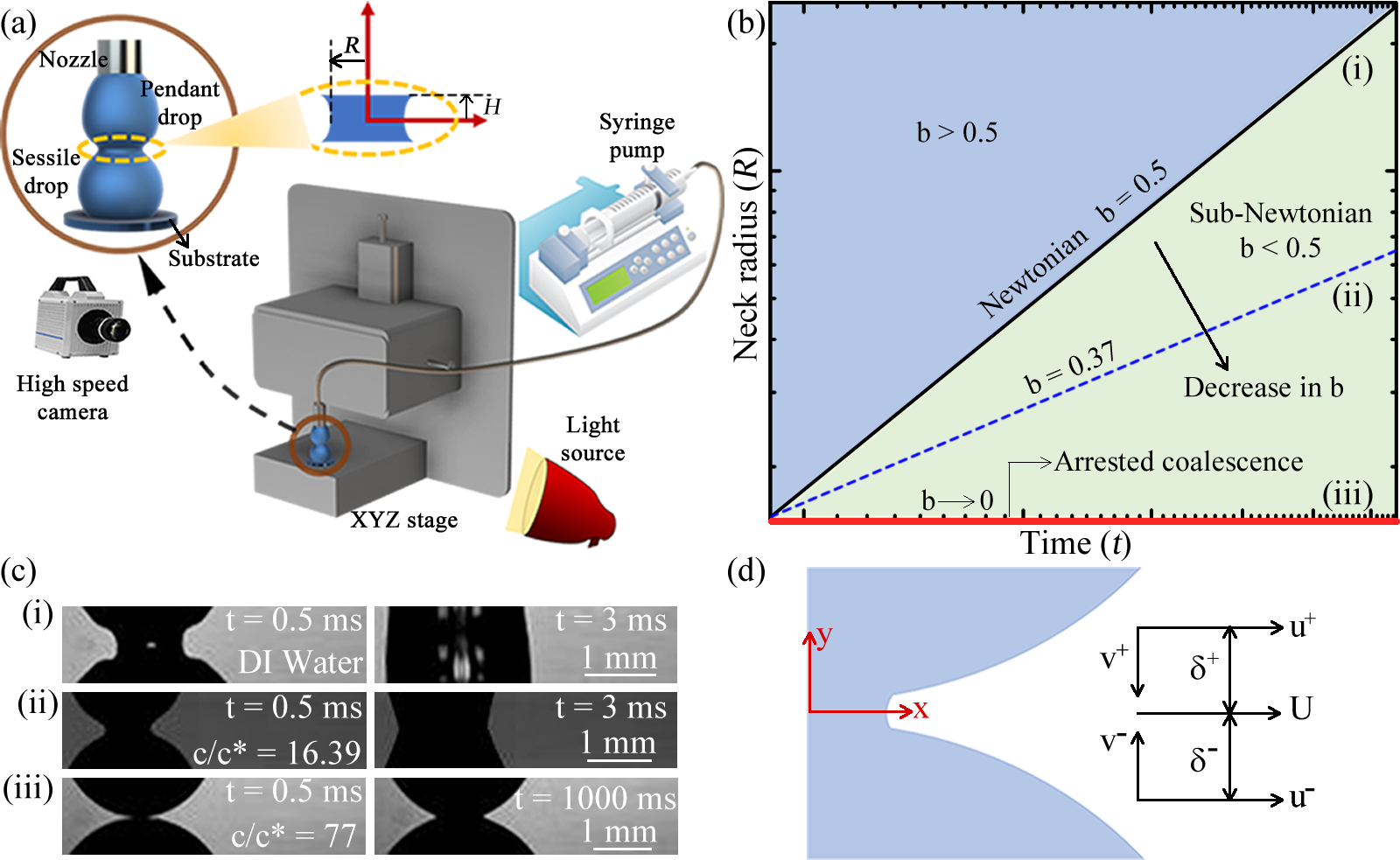}
\centering
\caption{(a) Schematic of experimental setup showing the geometric parameters neck radius $R$ and semi-bridge height $H$, (b) Regime map showing the classification of coalescence based on power-law exponent $b$. (c) Snapshots showing the neck at different time instants for (i) DI water, (ii) $c/c^*=16.39$ of PEO $M_w=5\times10^6$ g/mol and (iii) $c/c^*=77$ PEO $M_w=1\times10^5$ g/mol indicating the Newtonian, sub-Newtonian and arrested coalescence respectively, (d) Schematic of velocity distribution at the neck region for sessile-pendant coalescence.}
\label{fig:regimemap} 
\end{figure*}

Recent study by Chen et al.\cite{CHEN2022283} on a subclass of viscoelastic fluids i.e polymers and gels showed that at late time scales coalescence is slower than the Newtonian drops. This conclusion was drawn based on stress relaxation behaviour in polymers using molecular dynamics simulation.  Similarly Xu et al.\cite{xu2022bridge} also reported a slower growth of neck during coalescence of two immiscible Newtonian droplets. Another study, which employed numerical experiments on sessile-sessile drop coalescence of power-law fluids, showed a deviation from Newtonian behaviour as a function of the power-law exponent \cite{chen2022probing}. Our previous study on polymeric droplet coalescence \cite{varma2022rheocoalescence} highlighted the relevance of macromolecular relaxation time on the neck radius evolution $R$. The schematic of the neck during coalescence is shown in Fig.~\ref{fig:regimemap}(a). Instead of viscous versus inertial regime delineation in Newtonian fluids, aqueous solutions of macromolecules showed three different regimes namely, inertio-elastic, viscoelastic and elasticity dominated regimes. The non-dimensional parameter concentration ratio, $c/c^*$, governed the appearance of the various regimes. It has also been shown that the temporal evolution of the bridge follows a universal behaviour in inertio-elastic and viscoelastic regimes i.e $R\sim t^{b}$ (where, $b$ is power-law exponent) with $b=0.37$ along with continuously decreasing $b$ in the elasticity dominated regime. Based on scaling analysis using linear Phan-Thein-Tanner (PTT)\cite{ptt1,ptt2} constitutive equation the study proposed a time scale $\tau^*=\sqrt{\frac{\eta\lambda}{\rho R_o^2}}$ where $\eta$, $\rho$ and $R_o$ are the viscosity, density and length scale of the polymeric droplet to capture neck evolution exponent $b$. Whether these deviations from Newtonian behaviour in macromolecular fluids is due to visco-elasticity or shear dependent viscosity remains unresolved.

In the current study, we have developed a theoretical model to capture the neck radius evolution across different regimes. The theoretical framework is developed based on the unification of Oldroyd-B, linear PTT, Gesikus constitutive equations\cite{bird} under high Weissenberg number $Wi$ creeping flow limit. Thus this resolves the unanswered argument that visco-elasticity is the reason for deviation in neck evolution dynamics. However, for simplicity we use Giesekus constitutive equation\cite{giesekus1982simple} to explain the model. The proposed model is validated with the experiments performed with Poly(ethylene oxide) (PEO) for a range of molecular weights $M_w$. Apart from the theoretical model, we have also demonstrated the effect of entanglement density $N_e$ on coalescence to propose a new time scale $T^*$ that intertwines continuum and molecular approaches to macromolecular fluid. Using the  analogy of sub-diffusive behavior in Brownian dynamics, we name this broader class of sluggish merging of two droplets as sub-Newtonian coalescence. For the sessile-pendant droplet configuration, the regime of sub-Newtonian coalescence is marked by a power-law exponent (b) such that $0<b<0.5$. The right-hand limit corresponds to a purely viscous Newtonian fluid, whereas the left-hand limit corresponds to arrested-coalescence. 
Macromolecular fluids coalescence are such examples of sub-Newtonian coalescence where the presence of an additional resistance by elastic force slows down the coalescence resulting in an exponent $b$ lesser than the universal $b=0.5$ for Newtonian fluids. Even in Newtonian droplet coalescence, if the drops are immiscible one sees sub-Newtonian coalescence. Further we also propose a theoretical limit for an arrested coalescence ($b\rightarrow 0$). A regime map delineating the Newtonian ($b=0.5$), sub-Newtonian ($b<0.5$) and the arrested coalescence ($b\rightarrow 0$) is shown in Fig.~\ref{fig:regimemap}(b) with the blue dashed line representing the universality proposed for the polymeric fluids in our previous study\cite{our} and the solid red line signifying the arrested limit. Fig.~\ref{fig:regimemap}(c)represents the snapshots of the phenomenon at different time instants for DI water, $c/c^*=16.39$ of PEO $M_w=5\times10^6$ g/mol and $77$ of PEO $M_w=1\times10^5$ g/mol respectively. It can be observed from Fig.~\ref{fig:regimemap}(b) that the neck evolution slows down as one moves from from Newtonian to sub-Newtonian with the limiting case being the arrested coalescence. A detailed discussion on arrested coalescence is presented in the later sections. 

\subsection*{Theory}

To obtain the theoretical solution for coalescence phenomenon, we employ the symmetry of the problem and formulate our analysis in two dimensional Cartesian coordinates as shown in Fig.~\ref{fig:regimemap}(d). The kinematics of the flow is assumed to be quasi-radial at the neck region $y=0$ implying that $u\ne 0$ and $\mathrm{v}=0$. For a small region of width $\delta$ on either sides of $y=0$ line the flow field is such that $u^+=u^-\ne 0$ and $\mathrm{v}^+=-\mathrm{v}^-\ne 0$ respectively. Owing to the fact of mirror symmetry and quasi-radial assumption, the flow field at $y=0$ line has the constraints of $\frac{\partial u}{\partial y}=0$ and $\frac{\partial \mathrm{v}}{\partial x}= 0$.

The dynamics of the coalescence phenomenon is governed by conservation of mass and momentum equations along $y=0$ line as shown in Eq (2) and Eq (3a-b). To get stress tensor $\boldsymbol{\tau}$ in Eq (3) we employ to Giesekus constitutive equation represented in Eq (4). 
\begin{eqnarray}
\frac{\partial u}{\partial x}+\frac{\partial v}{\partial y}=0
\end{eqnarray}
\begin{subequations}
\begin{align}
\rho\big( u\frac{\partial u}{\partial x}\big)=-\frac{\partial p}{\partial x} +\frac{\partial \tau_{xx}}{\partial x}+\frac{\partial \tau_{xy}}{\partial y}\\
0=-\frac{\partial p}{\partial y}+\frac{\partial \tau_{yy}}{\partial y}
\end{align}
\end{subequations}
where, $\overset{\kern0.25em\smalltriangledown}{\boldsymbol{\tau}}=\frac{\partial\boldsymbol{\tau}}{\partial t}+\textbf{v}.\nabla{\boldsymbol{\tau}}-(\nabla \textbf{v})\boldsymbol{\tau}-\boldsymbol{\tau}(\nabla\textbf{v})^T$ is upper convected derivative.

By introducing the non-dimensional variables: $\textbf{v}^*=\textbf{v}/U$, $x^*=x/L$, $y^*=y/L$, $t^*=t/T$, $\boldsymbol{\tau}^*=\boldsymbol{\tau}/{\boldsymbol{\tau_c}}$, where $T:=L/U$, $U$, $L$, $\boldsymbol{\tau_c}$ are the characteristic time, velocity, length and stress respectively, Eq (4) is reduced to different forms as proposed in literature\cite{our,varma2022rheocoalescence}. Based on the dominant scale of $\boldsymbol{\tau_c}$ the constitutive law can be reduced to three distinct regimes, namely viscous dominant regime $\big(\boldsymbol{\tau_c}=\frac{\eta U}{L}\big)$, viscoelastic regime $\big(\boldsymbol{\tau_c}=Wi \frac{\eta U}{L}\big)$ where, $Wi=\frac{\lambda U}{L}$ is Weissenberg number and elasticity dominant regime $\big(\boldsymbol{\tau_c}=\frac{\eta}{\lambda}\big)$. Recently Varma et al.\cite{varma2022rheocoalescence} used the scale $\big(\boldsymbol{\tau_c}=\frac{\eta}{\lambda}\big)$ in elasticity dominated regime to predict the behaviour of $b$ during coalescence. For inertio-elastic and viscoelastic regimes a semi-analytical model is proposed by Varma et al.\cite{our} in which the upper convected derivative in Eq (4) was dropped. However, that approximation is not valid in the elasticity dominated regime as the Reynolds number $Re=\frac{\rho UL}{\eta_o}$ is $\mathcal{O}(10^{-5})$ and Weissenberg number $Wi$ is $\mathcal{O}(10^{3})$ as shown in our previous study\cite{varma2022elasticity} indicating that the flow has low $Re$ and high $Wi$. A detailed discussion on the $Re$ and $Wi$ for the present study is given in results and discussion section. For low $Re$ and high $Wi$ flows, the upper convective derivative of stress in the constitutive equation is the dominant term in Eq (3) (See supplementary information for the detailed derivation). Similarly, Renardy\cite{renardy1997high} showed that the upper convected derivative in the various constitutive equations\cite{bird} like Maxwell, Oldroyd-B, Linear PTT, exponential PTT model is the dominant term under quasi-steady assumption at high $Wi$ and low $Re$. As the upper convective derivative is the predominant term, the present theory is independent of continuum based constitutive equations.

\begin{eqnarray}
\tau_{c}\boldsymbol{\tau}^*+\frac{\tau_{c}\lambda U}{L}\big(\frac{\partial\boldsymbol{\tau}^*}{\partial t^*}+\overset{\kern0.25em\smalltriangledown}{\boldsymbol{\tau}}*+\frac{\alpha\tau_{c}L}{\eta U}\boldsymbol{\tau}^*\boldsymbol{\tau}^*\big)=2\frac{\eta U}{L}\textbf{D}^*
\end{eqnarray}
\begin{eqnarray}
\overset{\kern0.25em\smalltriangledown}{\boldsymbol{\tau}}*=2\textbf{D}^*.
\end{eqnarray}
As the coalescence phenomenon is predominately extensional, both the upper and lower convected derivative can be used to model the physics\cite{bellehumeur1998role}. In the present analysis, we have used the lower convected derivative ${\overset{\kern0.25em\smalltriangleup}{\boldsymbol{\tau}}}*=\frac{\partial\boldsymbol{\tau}}{\partial t}+\textbf{v}.\nabla{\boldsymbol{\tau}}+(\nabla \textbf{v})\boldsymbol{\tau}+\boldsymbol{\tau}(\nabla\textbf{v})^T$. For capturing the neck evolution dynamics, stress tensor in Eq (5) is simplified for the spatial region $y=0$ under the quasi-steady and quasi-radial assumptions to Eq (6a-c). The individual components of stress tensor are obtained by integrating Eq (6a-c) along $y=0$ line as represented in Eq (7a-c). 
\begin{subequations}
\begin{align}
u\frac{\partial \tau_{xx}}{\partial x} + 2\tau_{xx}\frac{\partial u}{\partial x}=2\frac{\eta}{\lambda}\frac{\partial u}{\partial x}\\
u\frac{\partial \tau_{yy}}{\partial x} -2\tau_{yy}\frac{\partial u}{\partial x}=-2\frac{\eta}{\lambda}\frac{\partial u}{\partial x}\\
u\frac{\partial \tau_{xy}}{\partial x}=0
\end{align}
\end{subequations}
\begin{subequations}
\begin{align}
\tau_{xx}=\frac{\eta}{\lambda} + H/u^2\\
\tau_{xy}=M\\
\tau_{yy}=\frac{\eta}{\lambda} + Fu^2
\end{align}
\end{subequations}
Here, $H,M$ and $F$ are integrating constants which in general are functions of $y$ locally.

To get the semi-analytical solution for the neck evolution, the momentum equation Eq (3) is further simplified to Eq (8) by introducing the scaling arguments $u\sim U$, $x\sim R$, $y\sim \frac{R^2}{2R_o}$, $\frac{\partial p}{\partial x}\sim\sigma\big(\frac{1}{R^2}+\frac{2R_o}{R^3}\big)$\cite{44} along with the components of stress tensor in Eq (3a-b) as $\frac{\partial\tau_{xx}}{\partial x}\sim\frac{\tau_{xx}}{R}$ and $\frac{\partial\tau_{xy}}{\partial y}\sim\frac{\tau_{xy}}{\frac{R^2}{2R_o}}$ ($R_o$ is droplet radius). Here, Eq (8) is a bi-quadratic equation of the form given in Eq (9) in which $A_1, A_2$ and $A_3$ are scaling constants. Eq (9) has 4 solutions in which two are negative and two are positive. However, the negative solutions are physically irrelevant as they suggest the neck collapses with time. Among the two acceptable solutions $\sqrt{\frac{P}{2}+\sqrt{\frac{P^2}{4}+Q}}$ captures the physical scenario. Therefore, the acceptable solution of Eq (9) is of the form Eq (10) where $U=\frac{dR}{dt}$.
\begin{eqnarray}
U^4-\bigg(\frac{A_1 \sigma}{\rho}\Big(\frac{1}{R}+\frac{2R_o}{R^2}\Big) + \frac{A_2}{\rho}\frac{\eta}{\lambda} + \frac{2A_3R_o}{\rho R}\bigg)U^2 -  \frac{A_2}{\rho}=0
\end{eqnarray}
\begin{eqnarray}
U^4 - P(\sigma,\rho, \eta,\lambda ,R, R_o,A_1,A_2,A_3)U^2 - Q(A_2,\rho)=0
\end{eqnarray}
\begin{eqnarray}
\frac{dR}{dt}= \sqrt{\frac{P}{2}+\sqrt{\frac{P^2}{4}+Q}}
\end{eqnarray}

Eq (10) is solved using first order finite difference scheme, in which time step $\Delta t$ is taken sufficiently small to ensure numerical stability. 

\subsection*{Results and Discussion}
Once the droplets touch each other, neck begins to grow for attaining the thermodynamic equilibrium state of a single daughter droplet. This neck growth is characterised by the temporal evolution of neck radius $R$ and semi-bridge height $H$ as represented in Fig.~\ref{fig:regimemap}(a). This evolution of neck radius at different time instants for concentration ratios $c/c^*$ 12.31 and 24.32 of $M_w=6\times10^5$ g/mol are shown in Fig. S1(a) and (b) respectively. Details of the experimental procedure is given in supplementary information. 

\begin{figure*}[h!]
\includegraphics[scale=0.31]{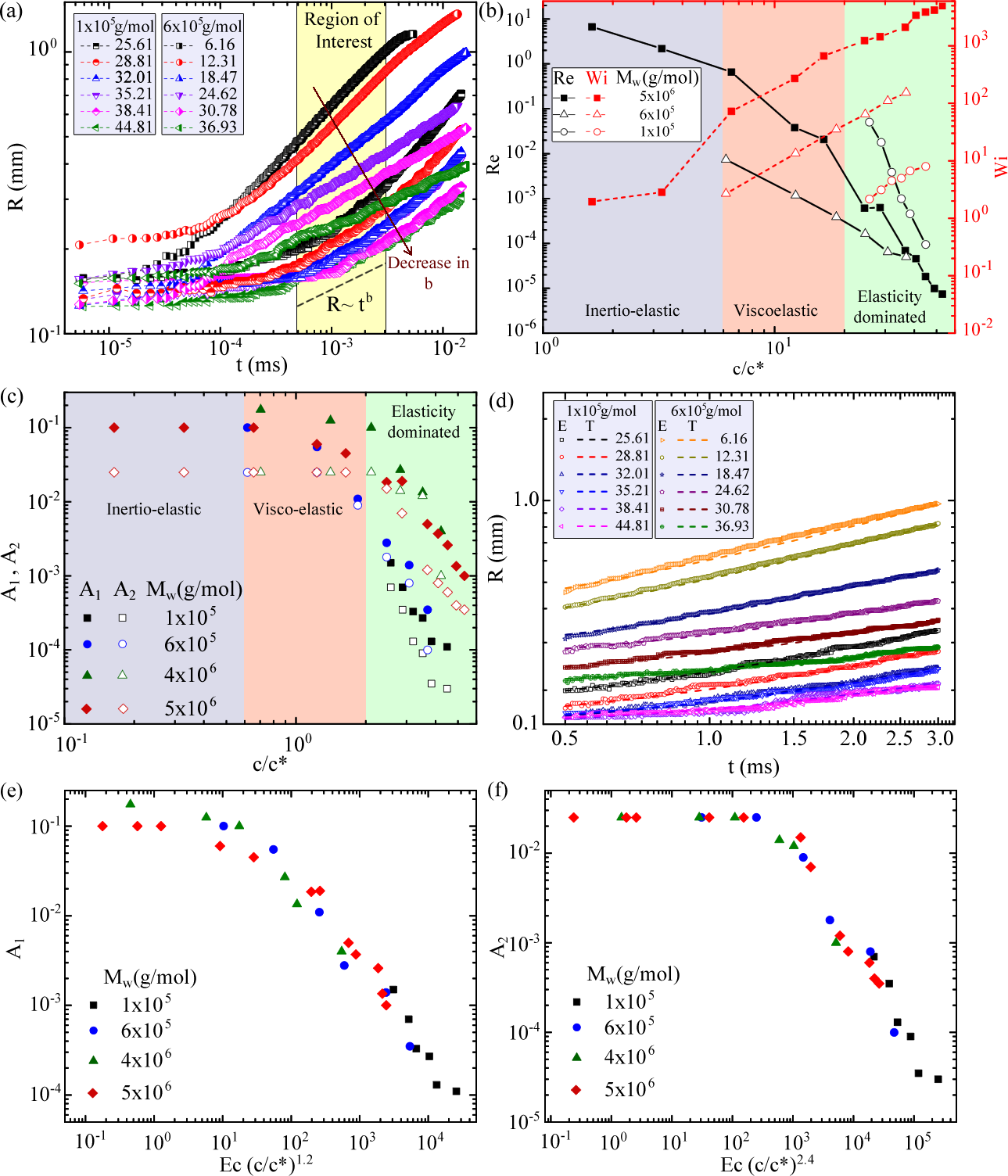}
\centering
\caption{(a) Growth in neck radius for various concentration ratios $c/c^*$ of PEO solutions showing the decrease in slope $b$ for $M_w=1\times10^5$ g/mol and $M_w=6\times10^5$ g/mol, (b) Regime map showing the variation of $Re$ and $Wi$ with $c/c^*$ across the molecular weight. (Note: Data for $M_w=5\times 10^6$ is obtained from our previous study\cite{varma2022elasticity}), (c) Dependence of scaling constants $A_1$ and $A_2$ on $c/c^*$ for various molecular weights across various regimes, namely, Inertio-elastic, visco-elastic and elasticity dominated regimes. (Note: $A_2$ values represented are the magnitudes), (d) Agreement between the experiments and solution of Eq (10) for various $c/c^*$ of $M_w=1\times10^5$ g/mol and $M_w=6\times10^5$ g/mol. Collapsing of scaling constants across molecular weights representing the dependence on $Ec$ for (e) $A_1$ and (f) $A_2$ in which $c/c^*$ is added empirically. (Note: $A_2$ values represented are the magnitudes).
}
\label{fig:raw} 
\end{figure*}

Neck radius evolution for various concentrations ratios of $M_w=1\times10^5$ and $6\times10^5$ g/mol is shown in Fig.~\ref{fig:raw}(a). The data represented for all the concentration ratios is of an average of 5 trials. It can be observed from Fig.~\ref{fig:raw}(a) that the neck radius evolution follows a power-law behaviour\cite{our} $R=at^b$ in the region of interest (ROI) along with decrease in power-law exponent $b$ with $c/c*$ which is consistent with our previous study\cite{varma2022rheocoalescence}. The error in measurement of $b$ is less than $\pm5\%$. 

To interpret the neck evolution during coalescence of polymeric droplets it is essential to outline the underlying forces. These are elastic force $F_e$, viscous force $F_v$, inertial force $F_i$ and capillary force $F_c$, where $F_c$ drives the phenomenon while the other three resist it. The relative magnitudes of these resistive forces are captured by Reynolds number $Re=<\frac{\rho UL}{\eta_o}>$, Weissenberg number $Wi=<\frac{\lambda U}{L}>$, Elasticity number $El=Wi/Re$ where $U\sim \partial R/\partial t$ and $L\sim R^2/{2R_o}$ are the characteristic scales associated with the flow. The variation of $Re$ and $Wi$ with $c/c^*$ is shown in Fig.~\ref{fig:raw}(b) with delineation based on the regimes proposed in our previous study\cite{varma2022rheocoalescence}(inertio-elastic (IE), viscoelastic (VE) and elasticity dominated (ED) regimes) for $M_w=1\times10^5$ and $M_w=6\times10^5$ g/mol along with the $M_w=5\times10^6$ g/mol. Fig.~\ref{fig:raw}(b) reveals that $Re<\mathcal{O}(10^{-2})$ and $Wi>\mathcal{O}(10^{0})$ for $M_w=1\times10^5$ and $M_w=6\times10^5$ g/mol indicating the flow has low $Re$ and high $Wi$.

In order to determine the closure to the analytical solution it is essential to study the physical behaviour of scaling parameters $A_1, A_2$ and $A_3$ represented in Eq (8). Among these scaling parameters, $A_1$ is the coefficient of capillary force that drives the coalescence dynamics while $A_2$ and $A_3$ are the coefficients of axial stress $\tau_{xx}$ and shear stress $\tau_{xy}$ respectively that oppose the neck growth. Owing to this nature, the coefficients $A_2$ and $A_3$ will have negative values while $A_1$ will have positive values. The magnitude of $A_1$ and $A_2$ used to get the numerical solution for all the chosen molecular weights with $c/c*$ are given in Fig.~\ref{fig:raw}(c). It is observed numerically that Eq (10) has least sensitivity to $A_3$, so we have assumed $A_3=A_2$ without loss of any generality. Fig.~\ref{fig:raw}(c) delineates the variation of $A_1$ and $A_2$ in IE, VE and ED regimes showing the constant values in IE and VE with a continuous decrease in ED regime. 

The neck evolution represented in ROI of Fig.~\ref{fig:raw}(a) is validated by numerically solving the proposed Eq (10) using the values of $A_1$ and $A_2$ given in Fig.~\ref{fig:raw}(c). Fig.~\ref{fig:raw}(d) shows the good agreement between the experiments and the proposed theory for $M_w=1\times10^5$ and $M_w=6\times10^5$ g/mol . To validate our theory further, the agreement between the present theory and the experimental results obtained from our previous study\cite{varma2022rheocoalescence} for $M_w=5\times10^6$ and $M_w=4\times10^6$ g/mol is shown in Fig. S2 (a) and (b). The power-law exponent $b$ obtained by fitting the experimental and the numerical data are given in Table-S1 (supplementary information) as $b$ and $b_{theo}$ respectively where $b_{theo}$ is power-law exponent obtained from the fitting of theoretical data. 

These scaling constants have a strong dependence on Elasto-capillary number $Ec=\frac{\text{Elastic Force}}{\text{Capillary Force}}=\frac{\eta_o R_o}{\sigma\lambda}$. By observing the dependence of $A_1$ and $A_2$ on $Ec$, $c/c*$ is added empirically in the form of $(c/c*)^{1.2}$ and $(c/c*)^{2.4}$ respectively to unify the functional form over different molecular weights as shown in Fig.~\ref{fig:raw}(e) and Fig.~\ref{fig:raw}(f) respectively. It is observed numerically that $b_{theo}$  has strong dependence on $A_1$. This is expected as $A_1$ is the coefficient of capillary forces that drive the coalescence dynamics. The parameter $A_1$ in Eq (8) corresponds to the relative contribution of capillary forces during the neck growth. It has been observed from of our previous experiments\cite{varma2022rheocoalescence} that, as the elasticity of droplets increases, the curvature of neck formed during coalescence changes, leading to an increase in capillary forces. Therefore, the term corresponding to $A_1$ in Eq (8) increases in magnitude. In order to maintain an overall balance in Eq (8), the coefficient $A_1$ decreases. As $A_1$ needs to account for the capillary forces, it has a strong functional dependence on $Ec$ which can be observed in Fig.~\ref{fig:raw}(e). It can be noted from Fig.~\ref{fig:raw}(e) that $A_1$ has a nearly constant value till $Ec(c/c^*)^{1.2}\sim O(10^{0})$ and then it decreases as the elastic forces increase. This is also the signature of the presence of three different regimes in coalescence where the exponent $b$ is constant in the IE/VE regime and a function of relaxation time $\lambda$ in the ED regime\cite{varma2022rheocoalescence}. Similarly the scaling parameters $A_2$ and $A_3$ are the coefficients of stress contributions in Eq (8) representing the relative contribution of elastic force and viscous force. As shown in Fig.~\ref{fig:raw}(f), these parameters are almost constant for the value of $Ec(c/c^*)^{2.4}\sim O(10^{2})$ followed by a strong decrease with increase in $Ec(c/c^*)^{2.4}$. It is further explained by Fig.~\ref{fig:raw}(f), as a transition in coalescence regime from IE/VE to ED regime. As we move towards the elasticity dominated regime, the value of $Ec(c/c^*)^{2.4}$ increases owing to higher elasticity of droplets. This increase in elasticity of droplets increases the stress contributions in Eq (8). Therefore, to retain the balance in equation, parameters $A_2$ and $A_3$ decrease monotonically.

\begin{figure*}[h!]
\includegraphics[width=\linewidth]{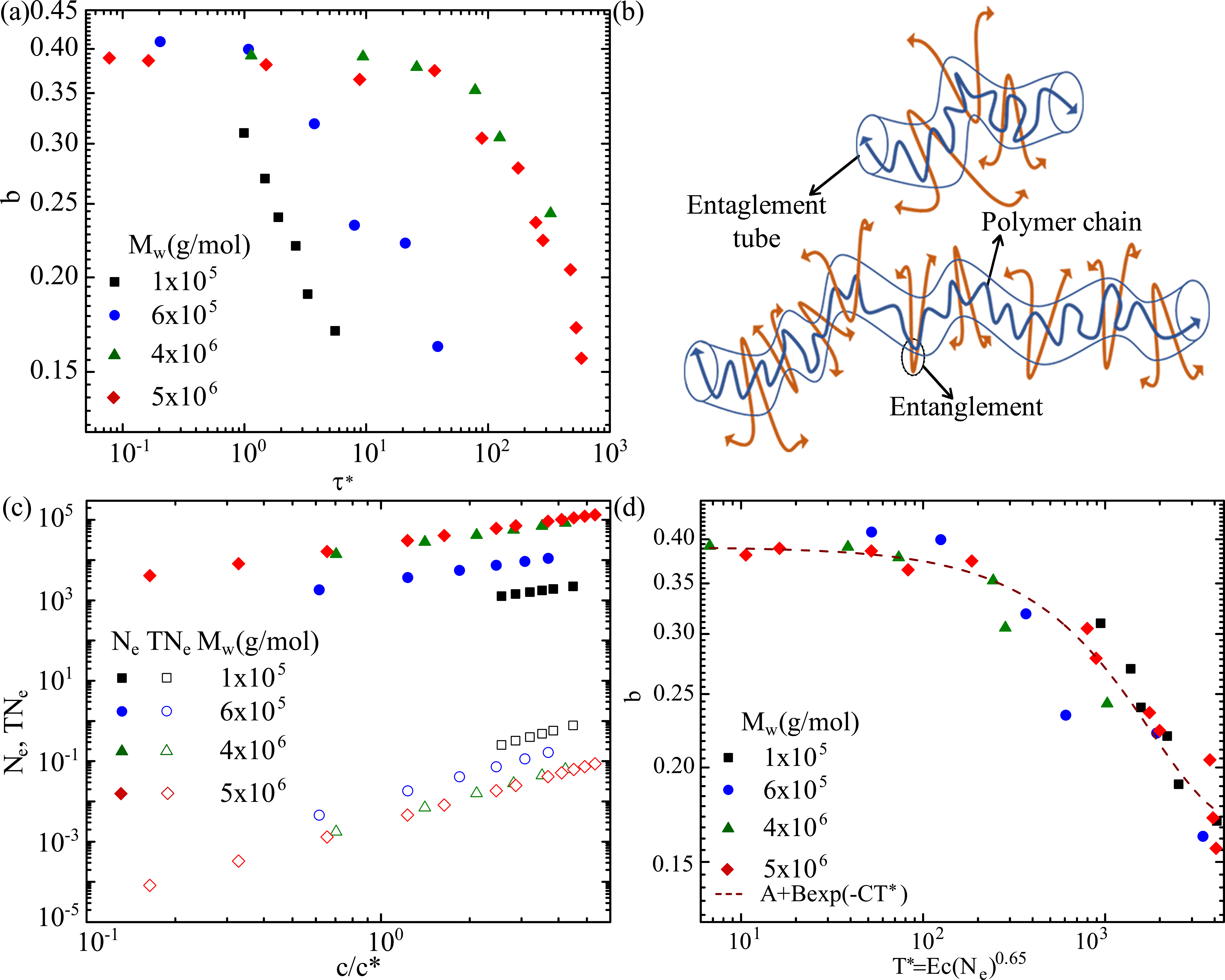}
\centering
\caption{(a) Variation of $b$ with the dimensionless time $\tau^*$ proposed in our previous study\cite{varma2022rheocoalescence} across different molecular weights, (b) Schematic of the chain entanglements for two different chain lengths, (c) Dependence of entanglement density $N_e$  and total entanglement density $TN_e=N_e\times N_d$ ($N_d$ is the order of number of chains) on $c/c^*$ across molecular weights and (d) Dependence of the power-law exponent $b$ on $T^*$ (ratio of $\tau_v$ and relaxation time $\lambda$) for various molecular weights. The dashed line represents the exponential fit of 94\% confidence interval with $A=0.1653\pm 0.01532$, $B=0.22545\pm 0.01508$, and $C=0.00076\pm 0.00017$.}
\label{fig:bvstau} 
\end{figure*}

The power-law exponent $b$ is the signature of the dominant governing force in coalescence dynamics. Apart from the early time scale where the exponent value is unity, it has a regime-dependent value for later time scales. For Newtonian, it is a universal value of $b=0.5$\cite{17,18,44}, while for polymeric droplets, it is $b\leq0.38$\cite{our}. This reduction in evolution exponent results from an additional resistance offered by elastic forces that slows the neck evolution thereby delaying coalescence. This sluggish merging of the two polymeric droplets owing to fluid elasticity is analogous to slowed diffusion of Brownian particles in viscoelastic media where diffusive exponent $\alpha$ less than unity in $<\Delta r^2>\propto t^{\alpha}$ marks the sub-diffusive regime\cite{sprakel2008brownian}. And therefore, polymeric droplet coalescence can equivalently be classified as an example of a border class of coalescence with $b<0.5$ which we name as the sub-Newtonian coalescence.  Our previous study on sub-Newtonian type coalescence (Rheocoalescence)\cite{varma2022rheocoalescence} showed, power-law exponent $b$ depends on $\tau^*=\sqrt{\frac{\eta_o\lambda}{\rho R_o^2}}$ (ratio of relaxation time and the Newtonian time scale). The corresponding dependence of $b$ for $M_w=6\times10^5$ and $M_w=1\times10^5$ g/mol along with the data obtained from our previous study\cite{varma2022rheocoalescence} for $M_w=5\times10^6$ g/mol and $M_w=4\times10^6$ g/mol is shown in Fig.~\ref{fig:bvstau}(a). It is observed that there is a significant deviation in $b$ vs $\tau^*$ for $M_w=6\times10^5$ and $M_w=1\times10^5$ g/mol compared to the other two molecular weights. This deviation is related to the macromolecular dynamics of the chains rather the continuum behaviour. Such dependence can be explained using the entanglement density i.e entanglment junctions per chain, $N_e=(\frac{M_w}{M_e})(\frac{c}{c^*})$\cite{huang2013concentrated}, where $M_e$ is entanglement molecular weight, which is 2000 g/mol for PEO\cite{nath2018dynamics}. The $N_e$ values for different $c/c^*$ and $M_w$ are given in Fig.~\ref{fig:bvstau}(c). It can be observed from Fig.~\ref{fig:bvstau}(c), the entanglement densities for same $c/c^*$ across the molecular weights differ by an order between $M_w=1\times10^5$, $6\times10^5$ g/mol and $5\times10^6$ g/mol. Whereas, for molecular weights of $4\times10^6$ g/mol and $5\times10^6$ g/mol, $N_e$ are of same order. As $N_e$ for $M_w=1\times10^5$ and $M_w=6\times10^5$ g/mol is less than the other two molecular weights, the chains have lesser topological constraints owing to the small chain lengths resulting in much lower relaxation times. The corresponding entanglement constrains for two different chain lengths is represented in Fig.~\ref{fig:bvstau}(b) as a schematic. This difference in relaxation times is reflected in the continuum approach through the shear modulus $G=\eta_o/\lambda$. The $G$ values corresponding to various $c/c^*$ and $M_w$ are given in Table-2. The difference in $G$ for same $c/c^*$ across the molecular weights suggests a stronger approach to solid like behaviour leading to a faster decrease in exponent $b$ for $M_w=1\times10^5$ and $M_w=6\times10^5$ g/mol. This is further understood by the scale of total entanglement density $TN_e=N_e\times N_d$ where $N_d$ is the scale of number of chains per unit volume. Fig.~\ref{fig:bvstau}(c) shows that for the same $c/c^*$, $TN_e$ increases as the $M_w$ decreases indicating higher density of entanglement junctions per unit volume for lower $M_w$ resulting in higher $G$ at same $c/c^*$. As the coalescence process is an energy minimization phenomenon, the increase in $G$ leads to increase in elastic energy per unit volume\cite{pawar2012arrested} $e=\frac{3}{2}G\epsilon^2$. Once this elastic energy predominates the surface energy, the coalescence is arrested. Such approach to arrested coalescence leads to faster decrease in $b$.

To account for the deviation of $b$ with $\tau^*$ across molecular weights as represented in Fig.~\ref{fig:bvstau}(a), we redefine the non-dimensional characteristic time as $T^*=\frac{\tau_v}{\lambda}(N_e)^{0.65}=Ec(N_e)^{0.65}$ (where, $\tau_v=\frac{\eta_o R_o}{\sigma}$). Here, $N_e^{0.65}$ is  added empirically.  This term accounts for the entanglement density which is the fingerprint of entanglement junctions in the continuum. Therefore, $T^*=Ec(N_e)^{0.65}$ is the corrected time scale to represent the behaviour of power-law exponent $b$. Such behaviour is represented in Fig.~\ref{fig:bvstau}(d). Addition of $N_e$ in $T^*$ suggests that the continuum approach is not complete until the molecular description of chain dynamics via entanglement densities are empirically added to the continuum description. Similarly, $c/c^*$ was empirically added in the functional dependence of the scaling constants $A_1$ and $A_2$ with $Ec$ to account for the entanglement densities. The collapse of data shown in Fig.~\ref{fig:bvstau}(d) follows an exponential decay function similar to that in our previous study\cite{varma2022rheocoalescence}.

Finally, we look at one of the limiting cases of Eq (10). For getting a physically acceptable solution, it is important to have non-negative sum under square-root. One such case that violates the condition is when $P=0$ as the term $Q$ is always negative. On substituting $P=0$, we obtain Eq (11) that can be further simplified to  obtain a cut-off radius in terms of material properties at which the coalescence is arrested. It is observed from Fig.~\ref{fig:raw}(c), the value of the term $|-\frac{A_1}{A_2}|$ is bounded by $\mathcal{O}(10)$. Under small angle limit $R=R_0\theta$ as represented in Fig. S3, Eq (11) is further simplified to obtain the $\theta_{arrest}$ (angle subtended by neck when coalescence is arrested) as represented in Eq (12).

\begin{eqnarray}
-\frac{A_1}{A_2} =  \frac{\frac{\eta}{\lambda}+\frac{2R_o}{R}}{\sigma(\frac{1}{R}+\frac{2R_o}{R^2})}
\end{eqnarray}

\begin{eqnarray}
\theta_{arrest} =  \frac{4.47}{\sqrt{Ec}}
\end{eqnarray}

The proposed value for $\theta_{arrest}$ is in good agreement with the value proposed in literature for high elasticity droplets\cite{ongenae2021activity}. It is important to note that our result over predicts the value which we owe to experimental sensitivity and the stronger assumption of $P=0$. To further validate $\theta_{arrest}$ given in Eq (12), experiments are performed on $c/c^*=77$ of $M_w=1\times 10^5$ g/mol having the properties $\eta_o=75$ Pa.s and $\lambda=17$ ms. In the current study, we have considered coalescence as arrested if $\dot{\gamma}_{arrest}<0.5\%(\dot{\gamma}_{DI water})$, where $\dot{\gamma}$ is shear rate (See supplementary information for shear rate calculations). The value of $\theta_{arrest}$ obtained from experiments is 0.35 radians which is in agreement with the value obtained from Eq (12) i.e. 0.48 radians. This experimentally obtained $\theta_{arrest}$ is in between the values obtained from Eq (12) and the relation proposed by Ongenae et al.\cite{ongenae2021activity} for the arrested coalescence. However, it is important to note that in sessile-pendant configuration, gravity becomes an important parameter at the higher time scales owing to which coalescence is no longer arrested.

\subsection*{Conclusion}

In the current work, we have developed a theoretical framework to model polymeric droplet coalescence. We have unified the various constitutive laws under high Weissenberg creep flow limit to obtain a scaling based neck evolution equation. The theoretical framework is validated across different molecular weights of Poly(ethylene oxide) (PEO) with experiments. Our experiments and theoretical model have both highlighted the importance of macromolecular parameters for understanding the coalescence dynamics. The study also reports an empirically corrected $T^*$ over our previous study to account for entanglement densities across different molecular weights. Theoretical framework is further validated by looking at a limiting case of arrested coalescence under small angle limit. The value we obtain for $\theta_{arrest}$ is found to be inversely proportional to $Ec^{1/2}$ and is validated with experiments along with the value proposed in literature. Finally, we name the coalescence as sub-Newtonian if $b<0.5$ with limiting case $b\rightarrow 0$ as arrested. However, the current framework implicitly assumes Weber Number $We=(\rho U^2 R_o)/\sigma \rightarrow 0$ and neglects the effect of surrounding fluids by assuming low approach velocity and air as the outer fluid respectively. Further studies on different complex fluids are required to broaden the class of sub-Newtonian coalescence along with the effect of higher approach velocities and different surrounding fluids. It also poses an open question about the existence of super-Newtonian coalescence $b>0.5$ where the merging dynamics will be driven by an additional force other than capillary  and therefore will hasten the coalescence dynamics.

\subsection*{Materials and methods}
Poly(ethylene oxide) (PEO) of molecular weight $M_w=6\times10^5$ and $1\times10^5$ g/mol, solutions of various concentrations $c$ are prepared by adding the sufficient quantity of polymer to DI Water. All the solutions are agitated at 300 rpm to ensure homogeneous dispersion. Concentrations are chosen such that the solutions are in semi-dilute entangled regime ($c>c_e$, where $c_e=6c^*$\cite{63} is an entanglement concentration and $c^*$ is the critical concentration). Critical concentration is obtained using the Mark-Houwink-Sakurada relation\cite{43} for PEO and Flory relation $\displaystyle c^*=\frac{1}{0.072M_w^{0.65}}$. For $M_w=4\times10^6$ and $5\times10^6$ g/mol data is taken from our previous study\cite{varma2022elasticity}. The corresponding values of critical concentration and entanglement concentration for the chosen polymers are given in Table-1. The concentrations and corresponding concentration ratios $c/c^*$ of the solutions are given in Table-S1 along with the rheology data in supplementary information.

\begin{table}
\centering
  \caption{List of molecular weights of polymers along with their critical and entanglement concentrations. (Note: * represents data obtained from Varma et al.\cite{varma2022elasticity})}
  \label{tbl:example}
  \begin{tabular}{llll}
    \hline
    Polymer  & $M_w$ (g/mol)  & $c^*$ (\% w/v)  & $c_e$ (\% w/v) \\
    \hline
    PEO   & $1\times10^5$   & 0.781   & 4.686   \\
    PEO   & $6\times10^5$   & 0.244  & 1.464   \\
    PEO   & $^*4\times10^6$   & 0.061  & 0.366   \\
    PEO   & $^*5\times10^6$   & 0.071  & 0.426   \\
    \hline
  \end{tabular}
\end{table}

Experiments are performed on a glass substrate coated with Polydimethylsiloxane (PDMS). Before coating the PDMS, substrates are cleansed with detergent followed by sonication with DI water and acetone for 20 mins each and later allowing them to dry in a hot air oven at 95$^\circ$C for 30 mins. PDMS and the curing agent (Syl Gard 184 Silicone Elastomer Kit, Dow Corning) are mixed in 1:10 ratio through agitation. This mixture is desiccated for 30 mins to remove the visible bubbles in the solution. Finally, the PDMS substrates are obtained by dripping the mixture on glass substrate and spin coating at 5000 rpm for 60 s.
\bibliography{achemso-demo}

\providecommand{\latin}[1]{#1}
\makeatletter
\providecommand{\doi}
  {\begingroup\let\do\@makeother\dospecials
  \catcode`\{=1 \catcode`\}=2 \doi@aux}
\providecommand{\doi@aux}[1]{\endgroup\texttt{#1}}
\makeatother
\providecommand*\mcitethebibliography{\thebibliography}
\csname @ifundefined\endcsname{endmcitethebibliography}
  {\let\endmcitethebibliography\endthebibliography}{}
\begin{mcitethebibliography}{33}
\providecommand*\natexlab[1]{#1}
\providecommand*\mciteSetBstSublistMode[1]{}
\providecommand*\mciteSetBstMaxWidthForm[2]{}
\providecommand*\mciteBstWouldAddEndPuncttrue
  {\def\EndOfBibitem{\unskip.}}
\providecommand*\mciteBstWouldAddEndPunctfalse
  {\let\EndOfBibitem\relax}
\providecommand*\mciteSetBstMidEndSepPunct[3]{}
\providecommand*\mciteSetBstSublistLabelBeginEnd[3]{}
\providecommand*\EndOfBibitem{}
\mciteSetBstSublistMode{f}
\mciteSetBstMaxWidthForm{subitem}{(\alph{mcitesubitemcount})}
\mciteSetBstSublistLabelBeginEnd
  {\mcitemaxwidthsubitemform\space}
  {\relax}
  {\relax}

\bibitem[Frenkel(1945)]{76}
Frenkel,~J. Viscous flow of crystalline bodies under the action of surface
  tension. \emph{J. phys.} \textbf{1945}, \emph{9}, 385\relax
\mciteBstWouldAddEndPuncttrue
\mciteSetBstMidEndSepPunct{\mcitedefaultmidpunct}
{\mcitedefaultendpunct}{\mcitedefaultseppunct}\relax
\EndOfBibitem
\bibitem[Villermaux and Bossa(2009)Villermaux, and Bossa]{1}
Villermaux,~E.; Bossa,~B. Single-drop fragmentation determines size
  distribution of raindrops. \emph{Nature Physics} \textbf{2009}, \emph{5},
  697\relax
\mciteBstWouldAddEndPuncttrue
\mciteSetBstMidEndSepPunct{\mcitedefaultmidpunct}
{\mcitedefaultendpunct}{\mcitedefaultseppunct}\relax
\EndOfBibitem
\bibitem[Pruppacher and Klett(2010)Pruppacher, and Klett]{3}
Pruppacher,~H.~R.; Klett,~J.~D. \emph{Microphysics of Clouds and
  Precipitation}; Springer, 2010; pp 10--73\relax
\mciteBstWouldAddEndPuncttrue
\mciteSetBstMidEndSepPunct{\mcitedefaultmidpunct}
{\mcitedefaultendpunct}{\mcitedefaultseppunct}\relax
\EndOfBibitem
\bibitem[Orme(1997)]{14}
Orme,~M. Experiments on droplet collisions, bounce, coalescence and disruption.
  \emph{Progress in Energy and Combustion Science} \textbf{1997}, \emph{23},
  65--79\relax
\mciteBstWouldAddEndPuncttrue
\mciteSetBstMidEndSepPunct{\mcitedefaultmidpunct}
{\mcitedefaultendpunct}{\mcitedefaultseppunct}\relax
\EndOfBibitem
\bibitem[Hopfes \latin{et~al.}(2021)Hopfes, Petersen, Wang, Giglmaier, and
  Adams]{HOPFES2021103723}
Hopfes,~T.; Petersen,~J.; Wang,~Z.; Giglmaier,~M.; Adams,~N. Secondary
  Atomization of Liquid Metal Droplets at Moderate Weber Numbers.
  \emph{International Journal of Multiphase Flow} \textbf{2021}, \emph{143},
  103723\relax
\mciteBstWouldAddEndPuncttrue
\mciteSetBstMidEndSepPunct{\mcitedefaultmidpunct}
{\mcitedefaultendpunct}{\mcitedefaultseppunct}\relax
\EndOfBibitem
\bibitem[Stewart and Mazza(2000)Stewart, and Mazza]{13}
Stewart,~S.; Mazza,~G. EFFECT OF FLAXSEED GUM ON QUALITY AND STABILITY OF A
  MODEL SALAD DRESSING 1. \emph{Journal of Food Quality} \textbf{2000},
  \emph{23}, 373--390\relax
\mciteBstWouldAddEndPuncttrue
\mciteSetBstMidEndSepPunct{\mcitedefaultmidpunct}
{\mcitedefaultendpunct}{\mcitedefaultseppunct}\relax
\EndOfBibitem
\bibitem[Ashgriz and Poo(1990)Ashgriz, and Poo]{5}
Ashgriz,~N.; Poo,~J. Coalescence and separation in binary collisions of liquid
  drops. \emph{Journal of Fluid Mechanics} \textbf{1990}, \emph{221},
  183--204\relax
\mciteBstWouldAddEndPuncttrue
\mciteSetBstMidEndSepPunct{\mcitedefaultmidpunct}
{\mcitedefaultendpunct}{\mcitedefaultseppunct}\relax
\EndOfBibitem
\bibitem[Djohari \latin{et~al.}(2009)Djohari, Mart{\'\i}nez-Herrera, and
  Derby]{6}
Djohari,~H.; Mart{\'\i}nez-Herrera,~J.~I.; Derby,~J.~J. Transport mechanisms
  and densification during sintering: I. Viscous flow versus vacancy diffusion.
  \emph{Chemical Engineering Science} \textbf{2009}, \emph{64},
  3799--3809\relax
\mciteBstWouldAddEndPuncttrue
\mciteSetBstMidEndSepPunct{\mcitedefaultmidpunct}
{\mcitedefaultendpunct}{\mcitedefaultseppunct}\relax
\EndOfBibitem
\bibitem[Ambrose \latin{et~al.}(2015)Ambrose, Livitz, Wessels, Kuhl, Lusche,
  Scherer, Voss, and Soll]{tumor}
Ambrose,~J.; Livitz,~M.; Wessels,~D.; Kuhl,~S.; Lusche,~D.~F.; Scherer,~A.;
  Voss,~E.; Soll,~D.~R. Mediated coalescence: a possible mechanism for tumor
  cellular heterogeneity. \emph{American journal of cancer research}
  \textbf{2015}, \emph{5}, 3485\relax
\mciteBstWouldAddEndPuncttrue
\mciteSetBstMidEndSepPunct{\mcitedefaultmidpunct}
{\mcitedefaultendpunct}{\mcitedefaultseppunct}\relax
\EndOfBibitem
\bibitem[Varma \latin{et~al.}(2020)Varma, Saha, Mukherjee, Bandopadhyay, Kumar,
  and Chakraborty]{our}
Varma,~S.~C.; Saha,~A.; Mukherjee,~S.; Bandopadhyay,~A.; Kumar,~A.;
  Chakraborty,~S. Universality in coalescence of polymeric fluids. \emph{Soft
  Matter} \textbf{2020}, \emph{16}, 10921--10927\relax
\mciteBstWouldAddEndPuncttrue
\mciteSetBstMidEndSepPunct{\mcitedefaultmidpunct}
{\mcitedefaultendpunct}{\mcitedefaultseppunct}\relax
\EndOfBibitem
\bibitem[Varma \latin{et~al.}(2021)Varma, Saha, and
  Kumar]{varma2021coalescence}
Varma,~S.~C.; Saha,~A.; Kumar,~A. Coalescence of polymeric sessile drops on a
  partially wettable substrate. \emph{Physics of Fluids} \textbf{2021},
  \emph{33}, 123101\relax
\mciteBstWouldAddEndPuncttrue
\mciteSetBstMidEndSepPunct{\mcitedefaultmidpunct}
{\mcitedefaultendpunct}{\mcitedefaultseppunct}\relax
\EndOfBibitem
\bibitem[Varma \latin{et~al.}(2022)Varma, Rajput, and
  Kumar]{varma2022rheocoalescence}
Varma,~S.~C.; Rajput,~A.~S.; Kumar,~A. Rheocoalescence: Relaxation time through
  coalescence of droplets. \emph{Macromolecules} \textbf{2022}, \relax
\mciteBstWouldAddEndPunctfalse
\mciteSetBstMidEndSepPunct{\mcitedefaultmidpunct}
{}{\mcitedefaultseppunct}\relax
\EndOfBibitem
\bibitem[Varma \latin{et~al.}(2022)Varma, Dasgupta, and
  Kumar]{varma2022elasticity}
Varma,~S.~C.; Dasgupta,~D.; Kumar,~A. Elasticity can affect droplet
  coalescence. \emph{arXiv preprint arXiv:2205.11815} \textbf{2022}, \relax
\mciteBstWouldAddEndPunctfalse
\mciteSetBstMidEndSepPunct{\mcitedefaultmidpunct}
{}{\mcitedefaultseppunct}\relax
\EndOfBibitem
\bibitem[Chen \latin{et~al.}(2022)Chen, Pirhadi, and Yong]{CHEN2022283}
Chen,~S.; Pirhadi,~E.; Yong,~X. Viscoelastic necking dynamics between
  attractive microgels. \emph{Journal of Colloid and Interface Science}
  \textbf{2022}, \emph{618}, 283--289\relax
\mciteBstWouldAddEndPuncttrue
\mciteSetBstMidEndSepPunct{\mcitedefaultmidpunct}
{\mcitedefaultendpunct}{\mcitedefaultseppunct}\relax
\EndOfBibitem
\bibitem[Xu \latin{et~al.}(2022)Xu, Wang, and Che]{xu2022bridge}
Xu,~H.; Wang,~T.; Che,~Z. Bridge evolution during the coalescence of immiscible
  droplets. \emph{Journal of Colloid and Interface Science} \textbf{2022},
  \relax
\mciteBstWouldAddEndPunctfalse
\mciteSetBstMidEndSepPunct{\mcitedefaultmidpunct}
{}{\mcitedefaultseppunct}\relax
\EndOfBibitem
\bibitem[Chen \latin{et~al.}(2022)Chen, Pan, Nie, Ma, Fang, and
  Yin]{chen2022probing}
Chen,~H.; Pan,~X.; Nie,~Q.; Ma,~Q.; Fang,~H.; Yin,~Z. Probing the coalescence
  of non-Newtonian droplets on a substrate. \emph{Physics of Fluids}
  \textbf{2022}, \emph{34}, 032109\relax
\mciteBstWouldAddEndPuncttrue
\mciteSetBstMidEndSepPunct{\mcitedefaultmidpunct}
{\mcitedefaultendpunct}{\mcitedefaultseppunct}\relax
\EndOfBibitem
\bibitem[Thien and Tanner(1977)Thien, and Tanner]{ptt1}
Thien,~N.~P.; Tanner,~R.~I. A new constitutive equation derived from network
  theory. \emph{Journal of Non-Newtonian Fluid Mechanics} \textbf{1977},
  \emph{2}, 353--365\relax
\mciteBstWouldAddEndPuncttrue
\mciteSetBstMidEndSepPunct{\mcitedefaultmidpunct}
{\mcitedefaultendpunct}{\mcitedefaultseppunct}\relax
\EndOfBibitem
\bibitem[Phan-Thien(1978)]{ptt2}
Phan-Thien,~N. A nonlinear network viscoelastic model. \emph{Journal of
  Rheology} \textbf{1978}, \emph{22}, 259--283\relax
\mciteBstWouldAddEndPuncttrue
\mciteSetBstMidEndSepPunct{\mcitedefaultmidpunct}
{\mcitedefaultendpunct}{\mcitedefaultseppunct}\relax
\EndOfBibitem
\bibitem[Bird \latin{et~al.}(1987)Bird, Armstrong, and Hassager]{bird}
Bird,~R.~B.; Armstrong,~R.~C.; Hassager,~O. \emph{Dynamics of polymeric
  liquids. Vol. 1, 2nd Ed. : Fluid mechanics}; Wiley, 1987\relax
\mciteBstWouldAddEndPuncttrue
\mciteSetBstMidEndSepPunct{\mcitedefaultmidpunct}
{\mcitedefaultendpunct}{\mcitedefaultseppunct}\relax
\EndOfBibitem
\bibitem[Giesekus(1982)]{giesekus1982simple}
Giesekus,~H. A simple constitutive equation for polymer fluids based on the
  concept of deformation-dependent tensorial mobility. \emph{Journal of
  Non-Newtonian Fluid Mechanics} \textbf{1982}, \emph{11}, 69--109\relax
\mciteBstWouldAddEndPuncttrue
\mciteSetBstMidEndSepPunct{\mcitedefaultmidpunct}
{\mcitedefaultendpunct}{\mcitedefaultseppunct}\relax
\EndOfBibitem
\bibitem[Renardy(1997)]{renardy1997high}
Renardy,~M. The high Weissenberg number limit of the UCM model and the Euler
  equations. \emph{Journal of non-newtonian fluid mechanics} \textbf{1997},
  \emph{69}, 293--301\relax
\mciteBstWouldAddEndPuncttrue
\mciteSetBstMidEndSepPunct{\mcitedefaultmidpunct}
{\mcitedefaultendpunct}{\mcitedefaultseppunct}\relax
\EndOfBibitem
\bibitem[Bellehumeur \latin{et~al.}(1998)Bellehumeur, Kontopoulou, and
  Vlachopoulos]{bellehumeur1998role}
Bellehumeur,~C.~T.; Kontopoulou,~M.; Vlachopoulos,~J. The role of
  viscoelasticity in polymer sintering. \emph{Rheologica acta} \textbf{1998},
  \emph{37}, 270--278\relax
\mciteBstWouldAddEndPuncttrue
\mciteSetBstMidEndSepPunct{\mcitedefaultmidpunct}
{\mcitedefaultendpunct}{\mcitedefaultseppunct}\relax
\EndOfBibitem
\bibitem[Xia \latin{et~al.}(2019)Xia, He, and Zhang]{44}
Xia,~X.; He,~C.; Zhang,~P. Universality in the viscous-to-inertial coalescence
  of liquid droplets. \emph{Proceedings of the National Academy of Sciences}
  \textbf{2019}, \emph{116}, 23467--23472\relax
\mciteBstWouldAddEndPuncttrue
\mciteSetBstMidEndSepPunct{\mcitedefaultmidpunct}
{\mcitedefaultendpunct}{\mcitedefaultseppunct}\relax
\EndOfBibitem
\bibitem[Wu \latin{et~al.}(2004)Wu, Cubaud, and Ho]{17}
Wu,~M.; Cubaud,~T.; Ho,~C.-M. Scaling law in liquid drop coalescence driven by
  surface tension. \emph{Physics of Fluids} \textbf{2004}, \emph{16},
  L51--L54\relax
\mciteBstWouldAddEndPuncttrue
\mciteSetBstMidEndSepPunct{\mcitedefaultmidpunct}
{\mcitedefaultendpunct}{\mcitedefaultseppunct}\relax
\EndOfBibitem
\bibitem[Paulsen \latin{et~al.}(2011)Paulsen, Burton, and Nagel]{18}
Paulsen,~J.~D.; Burton,~J.~C.; Nagel,~S.~R. Viscous to inertial crossover in
  liquid drop coalescence. \emph{Physical Review Letters} \textbf{2011},
  \emph{106}, 114501\relax
\mciteBstWouldAddEndPuncttrue
\mciteSetBstMidEndSepPunct{\mcitedefaultmidpunct}
{\mcitedefaultendpunct}{\mcitedefaultseppunct}\relax
\EndOfBibitem
\bibitem[Sprakel \latin{et~al.}(2008)Sprakel, van~der Gucht, Stuart, and
  Besseling]{sprakel2008brownian}
Sprakel,~J.; van~der Gucht,~J.; Stuart,~M. A.~C.; Besseling,~N.~A. Brownian
  particles in transient polymer networks. \emph{Physical Review E}
  \textbf{2008}, \emph{77}, 061502\relax
\mciteBstWouldAddEndPuncttrue
\mciteSetBstMidEndSepPunct{\mcitedefaultmidpunct}
{\mcitedefaultendpunct}{\mcitedefaultseppunct}\relax
\EndOfBibitem
\bibitem[Huang \latin{et~al.}(2013)Huang, Mednova, Rasmussen, Alvarez, Skov,
  Almdal, and Hassager]{huang2013concentrated}
Huang,~Q.; Mednova,~O.; Rasmussen,~H.~K.; Alvarez,~N.~J.; Skov,~A.~L.;
  Almdal,~K.; Hassager,~O. Concentrated polymer solutions are different from
  melts: Role of entanglement molecular weight. \emph{Macromolecules}
  \textbf{2013}, \emph{46}, 5026--5035\relax
\mciteBstWouldAddEndPuncttrue
\mciteSetBstMidEndSepPunct{\mcitedefaultmidpunct}
{\mcitedefaultendpunct}{\mcitedefaultseppunct}\relax
\EndOfBibitem
\bibitem[Nath \latin{et~al.}(2018)Nath, Mangal, Kohle, Choudhury, Narayanan,
  Wiesner, and Archer]{nath2018dynamics}
Nath,~P.; Mangal,~R.; Kohle,~F.; Choudhury,~S.; Narayanan,~S.; Wiesner,~U.;
  Archer,~L.~A. Dynamics of nanoparticles in entangled polymer solutions.
  \emph{Langmuir} \textbf{2018}, \emph{34}, 241--249\relax
\mciteBstWouldAddEndPuncttrue
\mciteSetBstMidEndSepPunct{\mcitedefaultmidpunct}
{\mcitedefaultendpunct}{\mcitedefaultseppunct}\relax
\EndOfBibitem
\bibitem[Pawar \latin{et~al.}(2012)Pawar, Caggioni, Hartel, and
  Spicer]{pawar2012arrested}
Pawar,~A.~B.; Caggioni,~M.; Hartel,~R.~W.; Spicer,~P.~T. Arrested coalescence
  of viscoelastic droplets with internal microstructure. \emph{Faraday
  discussions} \textbf{2012}, \emph{158}, 341--350\relax
\mciteBstWouldAddEndPuncttrue
\mciteSetBstMidEndSepPunct{\mcitedefaultmidpunct}
{\mcitedefaultendpunct}{\mcitedefaultseppunct}\relax
\EndOfBibitem
\bibitem[Ongenae \latin{et~al.}(2021)Ongenae, Cuvelier, Vangheel, Ramon, and
  Smeets]{ongenae2021activity}
Ongenae,~S.; Cuvelier,~M.; Vangheel,~J.; Ramon,~H.; Smeets,~B. Activity-induced
  fluidization and arrested coalescence in fusion of cellular aggregates.
  \emph{Frontiers in Physics} \textbf{2021}, 321\relax
\mciteBstWouldAddEndPuncttrue
\mciteSetBstMidEndSepPunct{\mcitedefaultmidpunct}
{\mcitedefaultendpunct}{\mcitedefaultseppunct}\relax
\EndOfBibitem
\bibitem[Arnolds \latin{et~al.}(2010)Arnolds, Buggisch, Sachsenheimer, and
  Willenbacher]{63}
Arnolds,~O.; Buggisch,~H.; Sachsenheimer,~D.; Willenbacher,~N. Capillary
  breakup extensional rheometry (CaBER) on semi-dilute and concentrated
  polyethyleneoxide (PEO) solutions. \emph{Rheologica Acta} \textbf{2010},
  \emph{49}, 1207--1217\relax
\mciteBstWouldAddEndPuncttrue
\mciteSetBstMidEndSepPunct{\mcitedefaultmidpunct}
{\mcitedefaultendpunct}{\mcitedefaultseppunct}\relax
\EndOfBibitem
\bibitem[Tirtaatmadja \latin{et~al.}(2006)Tirtaatmadja, McKinley, and
  Cooper-White]{43}
Tirtaatmadja,~V.; McKinley,~G.~H.; Cooper-White,~J.~J. Drop formation and
  breakup of low viscosity elastic fluids: Effects of molecular weight and
  concentration. \emph{Physics of Fluids} \textbf{2006}, \emph{18},
  043101\relax
\mciteBstWouldAddEndPuncttrue
\mciteSetBstMidEndSepPunct{\mcitedefaultmidpunct}
{\mcitedefaultendpunct}{\mcitedefaultseppunct}\relax
\EndOfBibitem
\end{mcitethebibliography}
\end{document}